\documentclass{elsart}
\usepackage{fleqn}

\usepackage{graphicx}
\usepackage[figuresright]{rotating}

\newcommand{\veezero}{\mbox{$V^{0}$}}

\newcommand{\lam}{\mbox{$\Lambda$}}
\newcommand{\lamb}{\mbox{$\overline{\Lambda}$}}

\newcommand{\ks}{\mbox{$K_{S}$}}

\newcommand{\pbar}{\mbox{$\overline{p}$}}

\newcommand{\llb}{\mbox{$\overline{p}p\rightarrow\overline{\Lambda}\Lambda$}}

\newcommand{\ls}{\mbox{$\overline{p}p \rightarrow \overline{\Sigma}^{0}
 \Lambda + c.c.$}}

\newcommand{\spsp}{\mbox{$\overline{p}p \rightarrow 
 \overline{\Sigma^{+}} \Sigma^{+}$}}
\newcommand{\smsm}{\mbox{$\overline{p}p \rightarrow 
 \overline{\Sigma^{-}} \Sigma^{-}$}}
\newcommand{\ksks}{\mbox{$\overline{p}p \rightarrow K_{S} K_{S}$}}

\newcommand{\ksksom}{\mbox{$\overline{p}p \rightarrow K_{S} K_{S} \omega$}}
\newcommand{\kskseta}{\mbox{$\overline{p}p \rightarrow K_{S} K_{S} \eta$}}
\newcommand{\kskspi}{\mbox{$\overline{p}p \rightarrow K_{S} K_{S} \pi^0$}}

\newcommand{\ksksa}{\mbox{$\overline{p} A \rightarrow K_{S} K_{S} X$}}

\newcommand{\AmS}{{\protect\the\textfont2
  A\kern-.1667em\lower.5ex\hbox{M}\kern-.125emS}}

\begin{document}
\begin{frontmatter}

% declarations for front matter
%%%%%%%%%%%%%%%%%%%%%%%%%%%%%%%%%%%%%%%%%%%%%%%%%%%%%%%%%%%%%%%%%%%%%%%%%%%%%%%%
%
\title{Measurement of the $\overline{p}p \rightarrow K_S K_S \eta$ cross
    section at beam momenta in the regions of 1.45 and 1.7~GeV/c}
%
%%%%%%%%%%%%%%%%%%%%%%%%%%%%%%%%%%%%%%%%%%%%%%%%%%%%%%%%%%%%%%%%%%%%%%%%%%%%%%%%

\author[CMU]     {P.D.~Barnes},
\author[Illinois]{B.~Bunker},
\author[Erlangen]{H.~Dennert},
\author[Illinois]{R.A.~Eisenstein},
\author[Erlangen]{W.~Eyrich},
\author[Freiburg]{H.~Fischer},
\author[CMU]     {G.~Franklin},
\author[Freiburg]{J.~Franz},
\author[Julich]  {R.~Geyer},
\author[Illinois]{P.~Harris},
\author[Erlangen]{J.~Hauffe},
\author[Illinois]{D.~Hertzog},
\author[Uppsala] {T.~Johansson},
\author[Illinois]{T.~Jones},
\author[Julich]  {K.~Kilian},
\author[Julich]  {W.~Oelert},
\author[Uppsala] {S.~Pomp\corauthref{cor}},
\ead{pomp@tsl.uu.se}
\author[CMU]     {B.~Quinn},
\author[Julich]  {K.~R\"{o}hrich},
\author[Freiburg]{E.~R\"{o}ssle},
\author[Julich]  {K.~Sachs},
\author[Freiburg]{H.~Schmitt},
\author[Julich]  {T.~Sefzick},
\author[CMU]     {J.~Seydoux},
\author[Erlangen]{F.~Stinzing},
\author[Illinois]{R.~Tayloe},
\author[Freiburg]{R.~Todenhagen},
\author[Uppsala] {E.~Traneus},
\author[Erlangen]{S.~Wirth}

%%%%%%%%%%%%%%%%%%%%%%%%%%%%%%%%%%%%%%%%%%%%%%%%%%%%%%%%%%%%%%%%%%%%%%%%%%%%%%%%

\corauth[cor]{Corresponding author.}

\address[CMU]{
  Carnegie Mellon University, Pittsburgh, PA 15213, USA
}
\address[Erlangen]{
  Universit\"{a}t Erlangen-N\"{u}rnberg, 91058 Erlangen, Germany
}

\address[Freiburg]{
  Universit\"{a}t Freiburg, 79104 Freiburg, Germany
}

\address[Julich]{
  IKP, Forschungszentrum J\"{u}lich, 52428 J\"{u}lich, Germany
}

\address[Illinois]{
  University of Illinois at Urbana-Champaign, Urbana, IL 61801, USA
}

\address[Uppsala]{
  Uppsala University, 75121 Uppsala, Sweden
}

%%%%%%%%%%%%%%%%%%%%%%%%%%%%%%%%%%%%%%%%%%%%%%%%%%%%%%%%%%%%%%%%%%%%%%%%%%%%%%%%
%        \thanks{Footnotes should appear on the first page only to
%                indicate your present address (if different from your
%                normal address), research grant, sponsoring agency, etc.
%                These are obtained with the {\tt\ttbs thanks} command.}
%        and
%        X.-Y. Wang\address{Economics Department, University of Winchester,\\
%        2 Finch Road, Winchester, Hampshire P3L T19, United Kingdom}}
%
%%%%%%%%%%%%%%%%%%%%%%%%%%%%%%%%%%%%%%%%%%%%%%%%%%%%%%%%%%%%%%%%%%%%%%%%%%%%%%%%
%\begin{document}
%%%%%%%%%%%%%%%%%%%%%%%%%%%%%%%%%%%%%%%%%%%%%%%%%%%%%%%%%%%%%%%%%%%%%%%%%%%%%%%%

%
% typeset front matter
%

%\maketitle

%%%%%%%%%%%%%%%%%%%%%%%%%%%%%%%%%%%%%%%%%%%%%%%%%%%%%%%%%%%%%%%%%%%%%%%%%%%%%%%%
\begin{abstract}
%%%%%%%%%%%%%%%%%%%%%%%%%%%%%%%%%%%%%%%%%%%%%%%%%%%%%%%%%%%%%%%%%%%%%%%%%%%%%%%%

The PS185 experiment at LEAR/CERN has investigated strangeness production
in $\overline{p}p$ collisions with final states such as
{\lamb\lam},
{$\overline{\Sigma}^0 \Lambda + c.c$},
{$\overline{\Sigma^+} \Sigma^+$},
{$\overline{\Sigma^-} \Sigma^-$} and
{\ks\ks}.
%%Results are now presented of a study of
Results are presented from a study of
about 32,000 {$\ks\ks X$} events obtained at
several $\overline{p}$ momenta in the
regions of 1.45 and 1.7~GeV/c.
The $\pbar p \rightarrow \ks \ks \eta$ cross sections extracted at
these momenta 
constitute the first measurement of this reaction in flight and
are broadly consistent with expectations of
a phase-space extrapolation of branching ratios from annihilation at rest.
\end{abstract}

\begin{keyword}
Eta meson \sep Strangeness production \sep LEAR
\PACS 25.43.+t \sep 13.75.Cs \sep 13.60.Le
\end{keyword}

\end{frontmatter}

%%%%%%%%%%%%%%%%%%%%%%%%%%%%%%%%%%%%%%%%%%%%%%%%%%%%%%%%%%%%%%%%%%%%%%%%%%%%%%%%
%%%%%%%%%%%%%%%%%%%%%%%%%%%%%%%%%%%%%%%%%%%%%%%%%%%%%%%%%%%%%%%%%%%%%%%%%%%%%%%%
%
                  \section{INTRODUCTION}
%
%%%%%%%%%%%%%%%%%%%%%%%%%%%%%%%%%%%%%%%%%%%%%%%%%%%%%%%%%%%%%%%%%%%%%%%%%%%%%%%%
%%%%%%%%%%%%%%%%%%%%%%%%%%%%%%%%%%%%%%%%%%%%%%%%%%%%%%%%%%%%%%%%%%%%%%%%%%%%%%%%

The PS185 experiment at the Low Energy Antiproton Ring, LEAR, at CERN
was designed to study strangeness production in $\pbar p$ collisions.
Most attention has been drawn to the two-body  {\llb}
reaction~\cite{lambda}.
Several observables, such as total and differential cross sections,
polarisations and spin correlations, have been measured
over the momentum region from threshold to 2~GeV/c.
PS185 has also investigated other two--body final states in the
{\ls}~\cite{sigma},
{\spsp} and {\smsm}~\cite{sig+-}
and {\ksks}~\cite{ksks} reactions.

These studies are now extended
to many--body final--states with two neutral strange particles.
%Results for the hyperon polarisation in {\llbx} are found in
%Refs.~\cite{Pomp99,Pomp00}.
We present in this paper a measurement of the
{\kskseta} cross section
from $\overline{p}p$ annihilation in flight. This
constitutes the first cross section measurement of the reaction since
the only other available data on this final state is
the branching ratio from $\overline{p}p$ annihilation at rest, obtained
in a bubble chamber experiment~\cite{Bara67}.

A study of the {\kskseta} reaction is interesting because all three
final--state particles contain strangeness and this may provide
additional information on strangeness production. In a constituent quark
model, the process can be viewed
as the annihilation of quark--antiquark pairs with the subsequent
creation of strange--antistrange quark pairs. Such models are depicted in
Fig.~1a. The process could also
proceed via the dissociation of a strangeness component in the
(anti-)proton, as shown in Fig.~1b, an idea which has already been applied
to the {\llb} reaction~\cite{Alb95}.
More conventionally, the reaction can be described in a one-baryon-exchange
model via charged $\Sigma$ exchange. The final state can then either be
directly produced (Fig.~1c) or proceed via the excitation
of a resonance (Fig.~1d).
In the latter case one would expect the $K^*_3(1780)$
resonance to be important in flight since it is the only known $K^*$ resonance in this
mass region with a substantial decay branching ratio into the $\ks \eta$
channel
$(30 \pm 13)\%$~\cite{pdg2000}.
However, to the best of our knowledge,
no theoretical calculation has been made for this reaction.

%%%%%%%%%%%%%%%%%%%%%%%%%%%%%%%%%%%%%%%%%%%%%%%%%%%%%%%%%%%%%%%%%%%%%%%%%%%%%%%%
%
%                  figure quarklines
%
%%%%%%%%%%%%%%%%%%%%%%%%%%%%%%%%%%%%%%%%%%%%%%%%%%%%%%%%%%%%%%%%%%%%%%%%%%%%%%%%
\begin{figure}[hbt!]
\vspace{-0.3cm}
\begin{center}
 \resizebox{0.9\textwidth}{!}{
 \includegraphics*[2.2cm,16.4cm][17.5cm,25.5cm]
        {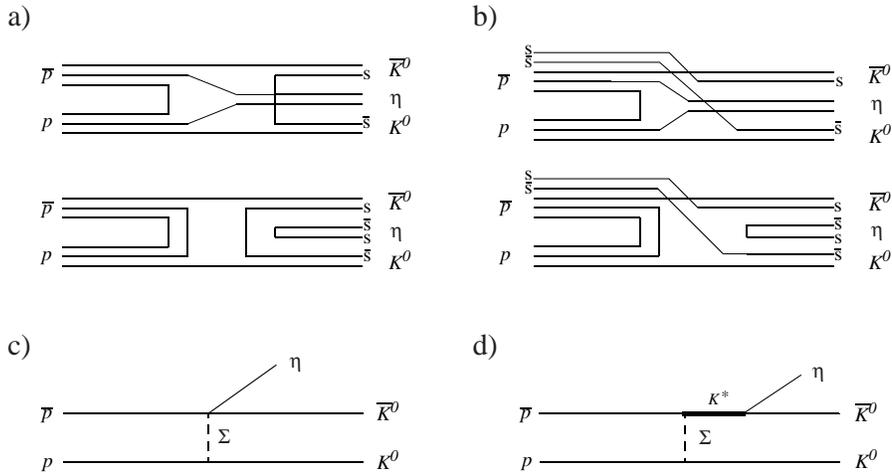}}
 \caption{
   Examples of diagrams for $\ks \ks \eta$ production.
   The upper row shows quark-line diagrams
   without (a) and with (b) strangeness from the (anti-)proton;
   the lower row shows diagrams for one-baryon-exchange models (see text).
  }
 \label{fig:quarkline}
\end{center}
\end{figure}
%

%%%%%%%%%%%%%%%%%%%%%%%%%%%%%%%%%%%%%%%%%%%%%%%%%%%%%%%%%%%%%%%%%%%%%%%%%%%%%%%%
%%%%%%%%%%%%%%%%%%%%%%%%%%%%%%%%%%%%%%%%%%%%%%%%%%%%%%%%%%%%%%%%%%%%%%%%%%%%%%%%
%
               \section{EXPERIMENT AND ANALYSIS}
%
%%%%%%%%%%%%%%%%%%%%%%%%%%%%%%%%%%%%%%%%%%%%%%%%%%%%%%%%%%%%%%%%%%%%%%%%%%%%%%%%
%%%%%%%%%%%%%%%%%%%%%%%%%%%%%%%%%%%%%%%%%%%%%%%%%%%%%%%%%%%%%%%%%%%%%%%%%%%%%%%%

In this study, we used data from two PS185 runs.
The first involved six beam momentum settings
between 1.440 and 1.477~GeV/c, whereas
the second was performed at beam momenta of
1.657, 1.662 and 1.774~GeV/c.
We will use
$1.455\pm0.020$~GeV/c and
$1.715\pm0.060$~GeV/c as momenta for the data sets.

\begin{figure}[hbt!]
\vspace{-0.3cm}
\begin{center}
 \resizebox{0.72\textwidth}{!}{
 \includegraphics*[1.0cm,5.5cm][20cm,23.5cm]
        {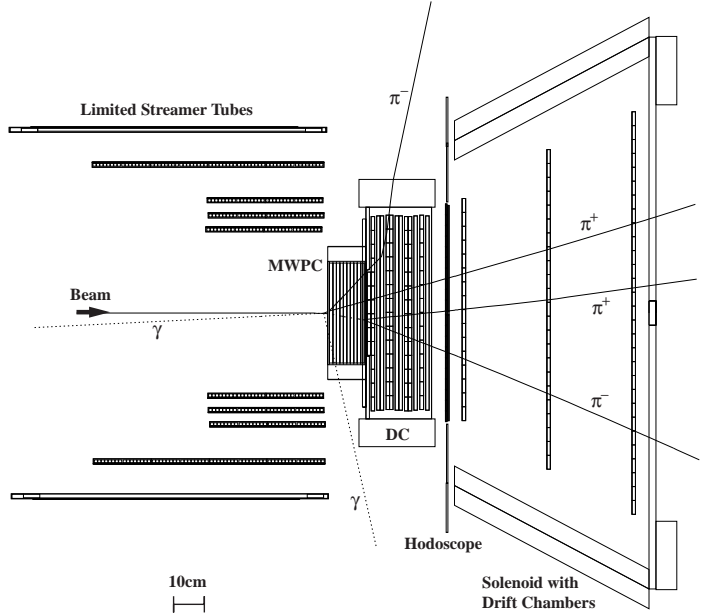}}
% \includegraphics*[0cm,0cm][21cm,19cm]
%        {figure2.ps}}
 \caption{
   The PS185 detector with a superimposed {\kskseta} event from the Monte
   Carlo simulation.
   The target is located just in front of the MWPC.
   The two $\gamma$'s originating from the $\eta$ decay are not detected.
  }
 \label{fig:det}
\end{center}
\end{figure}

The PS185 experimental set-up is shown in Fig.~\ref{fig:det}.
It employs a segmented target consisting of five
cylindrical cells each with a length and diameter of 2.5~mm.
The first cell is made from pure carbon
whereas the others consist of
polyethylene (CH$_{2}$).
The 5 cells are embedded and separated by scintillators
which are
used as vetoes to allow for triggering on events with neutral final
particles.
A two-layer scintillator hodoscope detects the charged decay products.
The sequence of
charged $\rightarrow$ neutral $\rightarrow$ charged states forms
the signature of the desired events and is the basis for
the triggering.

A chamber stack, consisting of multiwire proportional chambers (MWPC) and
drift chambers (DC), is located
between the target and the hodoscope.
A set of three drift chamber planes, contained
inside a magnetic field and placed downstream of the hodoscope,
allows a charge determination.
The backward region is covered by an arrangement of limited streamer tubes.

%%%%%%%%%%%%%%%%%%%%%%%%%%%%%%%%%%%%%%%%%%%%%%%%%%%%%%%%%%%%%%%%%%%%%%%%%%%%%%%%
%
%                  figure: detector with target
%
%%%%%%%%%%%%%%%%%%%%%%%%%%%%%%%%%%%%%%%%%%%%%%%%%%%%%%%%%%%%%%%%%%%%%%%%%%%%%%%%

Charged particle tracks are reconstructed in the chamber stack and combined
pairwise to form neutral decay {\veezero}'s.
A kinematical fit is then performed on the {\veezero}'s assuming that they
correspond to $\ks \rightarrow \pi^+ \pi^-$ decays.
It is based on a procedure described in Ref.~\cite{Frod79},
which has been extended to handle the case where
only angles and no momenta
are measured (cf.\ Ref.~\cite{Pomp99}).\
Monte Carlo simulations show that the  resulting uncertainty in the
reconstructed momenta is about 10\%.

%%%%%%%%%%%%%%%%%%%%%%%%%%%%%%%%%%%%%%%%%%%%%%%%%%%%%%%%%%%%%%%%%%%%%%%%%%%%%%%%
%
%                  figures wave and eloss
%
%%%%%%%%%%%%%%%%%%%%%%%%%%%%%%%%%%%%%%%%%%%%%%%%%%%%%%%%%%%%%%%%%%%%%%%%%%%%%%%%
\begin{figure}[hbt!]
\vspace{-0.3cm}
\begin{center}
 \resizebox{0.9\textwidth}{!}{
 \includegraphics*[1.5cm,0.0cm][19.5cm,18.0cm]
        {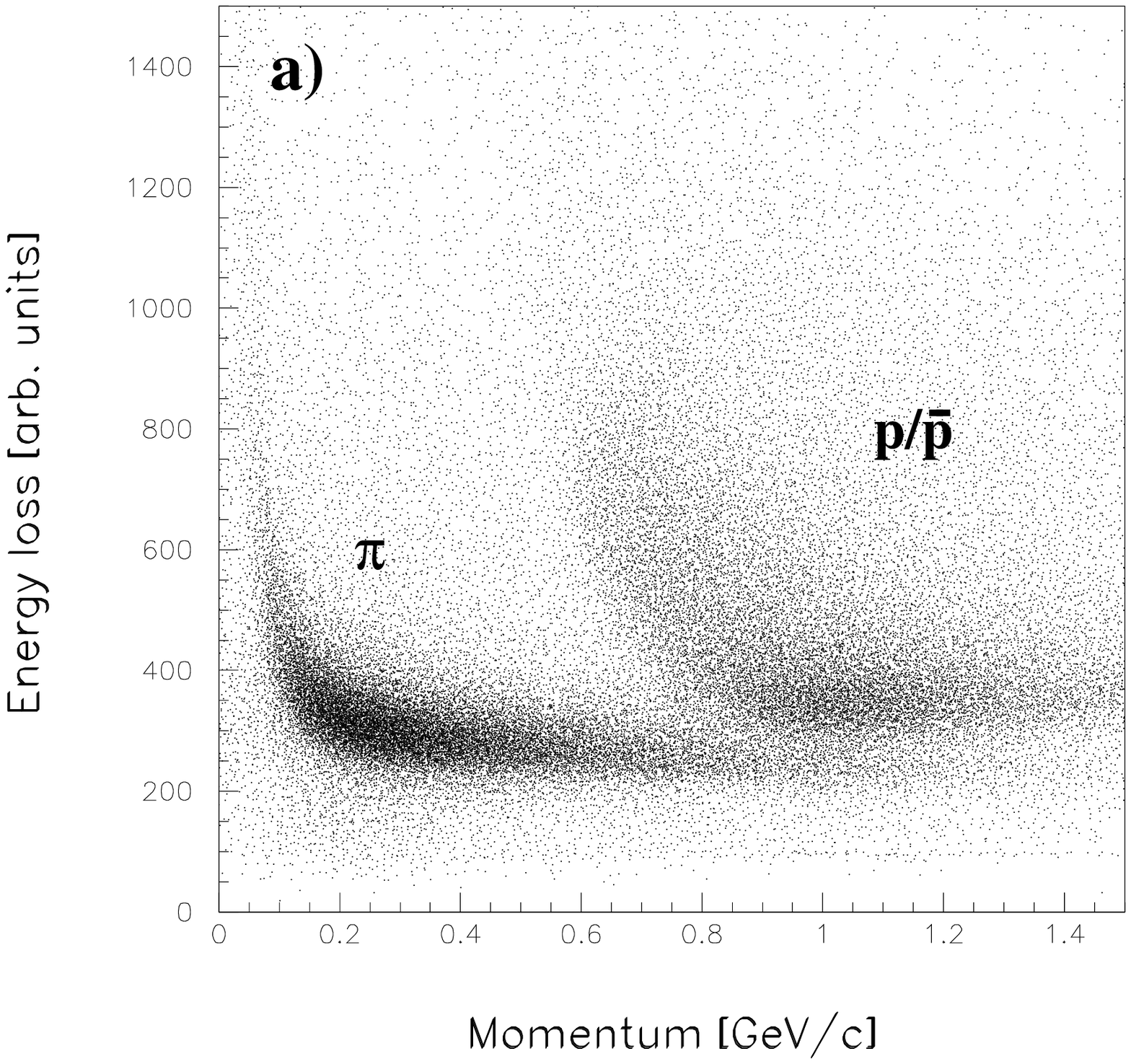}
 \includegraphics*[2.5cm,0.0cm][20.5cm,19.0cm]
        {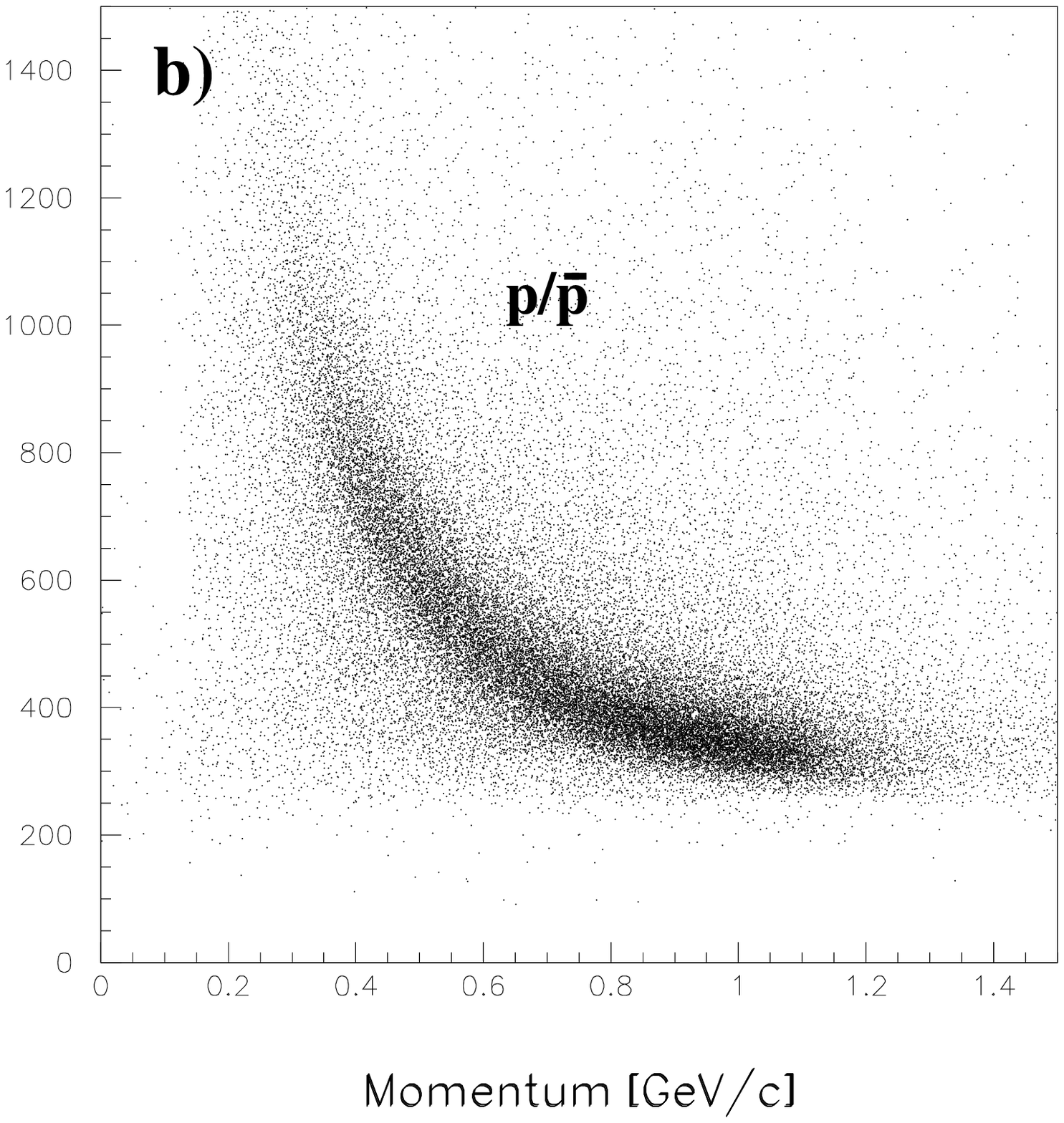}}
 \caption{
   (a) Energy losses from particle tracks in the hodoscope plotted versus their
   momenta for assumed $\ks$ decays.
   (b) Energy losses of assigned (anti-)proton tracks in the hodoscope plotted
   versus their momenta for
   assumed $\Lambda$($\overline{\Lambda}$) decays.
  }
 \label{fig:4_7-1990}
\end{center}
\end{figure}
%%%%%%%%%%%%%%%%%%%%%%%%%%%%%%%%%%%%%%%%%%%%%%%%%%%%%%%%%%%%%%%%%%%%%%%%%%%%%%%%

Both the $\ks\ks X$ and the $\Lambda\overline{\Lambda} X$ reaction channels
lead to events with two {\veezero}'s in the final state and it is not possible
to separate these channels using the kinematical fit alone since no momenta are
measured. However, the information from the energy loss in the hodoscope can
be used to differentiate between them.
Fig.~\ref{fig:4_7-1990}a shows a scatter plot of
the energy losses of particles in the hodoscope
versus their momenta obtained from the kinematical fit assuming a $\ks$ decay.
Bands belonging to pions and (anti-)protons are clearly visible.
The protons originate from $\Lambda$($\overline{\Lambda}$) decays
and constitute background events.
As a consistency check, these background events are refitted assuming
$\Lambda$($\overline{\Lambda}$) decays.
The results given in Fig.~\ref{fig:4_7-1990}b confirm the (anti-)proton
assignment for these events.

We obtained a total sample of 31,732 {$\ks\ks X$} events
originating from either $\pbar p$ or $\pbar$C interactions from the two
data-sets. Due to the relatively sharp lower edge of the Landau
distribution, the contamination in the event sample arising from the
hyperon background is estimated to be less than 2\%.
As a check of the final event sample, we determined
the lifetime for the assigned kaons.
Based on the measured decay length and the fitted momentum, we derived a
value of $(93 \pm 4)$~ps, which is in good agreement with the
PDG value of 89.35~ps~\cite{pdg2000}.

\begin{figure}[hbt!]
\begin{center}
 \resizebox{0.9\textwidth}{!}
 {\includegraphics*[0cm,7.5cm][19.5cm,25cm]{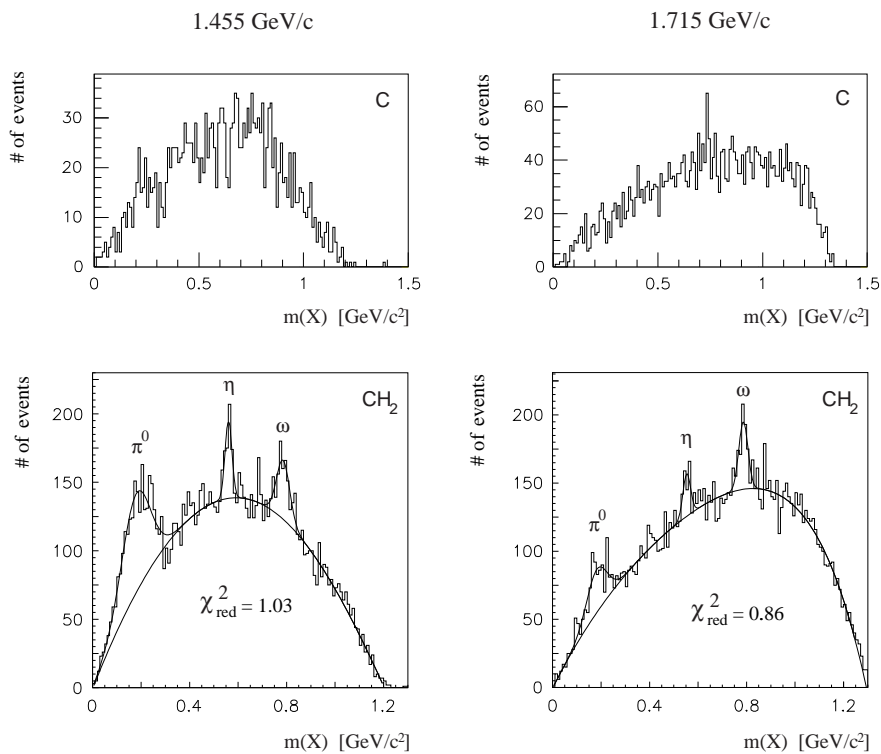}}
 \caption{
   Missing mass distributions for {\ksksa} events for the
   1.455~GeV/c (left) and 1.715~GeV/c (right) data-set.
   The upper row contains events from the carbon cell and
   the lower row events from the CH$_{2}$ cells.
   Peaks at the masses of the $\pi^0$, $\eta$ and $\omega$ mesons
   are visible in the spectra from the CH$_2$ cells.
   The solid lines represent fits
   assuming a 5th order polynomial and three gaussians.
  }
 \label{fig:ksksmissmass}
\end{center}
\end{figure}

Fig.~\ref{fig:ksksmissmass} illustrates the missing mass distributions
for the extracted {\ksksa} events. They show a broad
distribution stretching from 0 to 1.2 and 1.4~GeV/c$^2$,
respectively, which is due to $\ks \ks X$ production on $^{12}$C,
involving the production of $K^*$'s and $\pi^0$'s, as well as
combinatorial background. In addition, three peaks are visible in the
spectra from the CH$_2$ cells at the masses of the $\pi^0$, $\eta$ and
$\omega$ mesons (lower row) which are not found in the spectra from the
carbon cell (upper row). These peaks are attributed to the
{\kskspi}, {\kskseta} and {\ksksom} reactions.

%%%%%%%%%%%%%%%%%%%%%%%%%%%%%%%%%%%%%%%%%%%%%%%%%%%%%%%%%%%%%%%%%%%%%%%%%%%%%%%%
%
%                            miss mass figure
%
%%%%%%%%%%%%%%%%%%%%%%%%%%%%%%%%%%%%%%%%%%%%%%%%%%%%%%%%%%%%%%%%%%%%%%%%%%%%%%%%
%

%%%%%%%%%%%%%%%%%%%%%%%%%%%%%%%%%%%%%%%%%%%%%%%%%%%%%%%%%%%%%%%%%%%%%%%%%%%%%%%%
%%%%%%%%%%%%%%%%%%%%%%%%%%%%%%%%%%%%%%%%%%%%%%%%%%%%%%%%%%%%%%%%%%%%%%%%%%%%%%%%
%
              \section{RESULTS AND CONCLUSIONS}
%
%%%%%%%%%%%%%%%%%%%%%%%%%%%%%%%%%%%%%%%%%%%%%%%%%%%%%%%%%%%%%%%%%%%%%%%%%%%%%%%%
%%%%%%%%%%%%%%%%%%%%%%%%%%%%%%%%%%%%%%%%%%%%%%%%%%%%%%%%%%%%%%%%%%%%%%%%%%%%%%%%

%%%%%%%%%%%%%%%%%%%%%%%%%%%%
%\subsection{Number of events}
%%%%%%%%%%%%%%%%%%%%%%%%%%%%

We have made fits to the missing mass distributions from the CH$_2$ cells
to obtain the number of events with $\pi^0$'s, $\eta$'s and $\omega$'s in
the
final states arising from $\pbar p$ interactions. These involve a 5th order
polynomial for the background and three gaussians;
the results for the two data-sets are shown in Fig.~\ref{fig:ksksmissmass}.

The central values of the $\eta$ and $\omega$ peak are found to be
$(560 \pm 14)$~MeV/c$^2$ and $(786 \pm 27)$~MeV/c$^2$ for the 1.455~GeV/c
data and
$(553 \pm 17)$~MeV/c$^2$ and $(786 \pm 19)$~MeV/c$^2$ for the 1.715~GeV/c
data.
The ratios of the number of $\eta$ to $\omega$ events are
$0.691 \pm 0.139 \pm 0.069$ (1.455~GeV/c) and
$0.560 \pm 0.160 \pm 0.056$ (1.715~GeV/c),
where the first error is statistical and the second systematical.
The latter, which is due to the uncertainty in the background subtraction,
is estimated to be about 10\%. The $\pi^0$ case requires a slightly different
treatment which will be described later.

%%%%%%%%%%%%%%%%%%%%%%
%\subsection{Acceptance}
%%%%%%%%%%%%%%%%%%%%%%

The detector has the highest efficiency for forward-going (charged) particles.
Due to the influence of the $\eta$/$\omega$ mass difference on
the kinematics, we expect a larger acceptance for
{\kskseta} than for {\ksksom} events. This difference is less important
at higher energies.

The acceptances have been estimated by generating events with
phase space distributions and applying cuts representing the geometrical
acceptance and reconstruction efficiency. From these considerations we
obtain relative efficiencies
$\epsilon(\omega)/\epsilon(\eta) = 0.791 \pm 0.040$
at 1.455~GeV/c and $0.882 \pm 0.044$ at 1.715~GeV/c.
These results are consistent with the expectations mentioned above.

Different sets of cuts were tried and
it was found that the relative efficiencies are not very sensitive to
the specific set.
This is not unexpected since the topologies for the two types
of events are very similar provided that the differential distributions
of the two reactions are rather alike. We investigated this by
looking at the distribution of the c.m.~scattering angles.
We defined six regions in the missing mass spectrum: one covering each
of the $\eta$ and $\omega$ peaks and one on either side of them.
The c.m.~scattering angle distributions for these regions are
not significantly different from each other.
Furthermore, the experimental momentum and scattering angle distributions
and those
obtained from Monte Carlo simulations based on phase space distributions
are quite similar.

%%%%%%%%%%%%%%%%%%%%%%%%%
%\subsection{Normalisation}
%%%%%%%%%%%%%%%%%%%%%%%%%

\begin{figure}[htp!]
\vspace{-0.3cm}
\begin{center}
 \resizebox{0.52\textwidth}{!}{
 \includegraphics*[2.5cm,7.0cm][18.0cm,24.0cm]
        {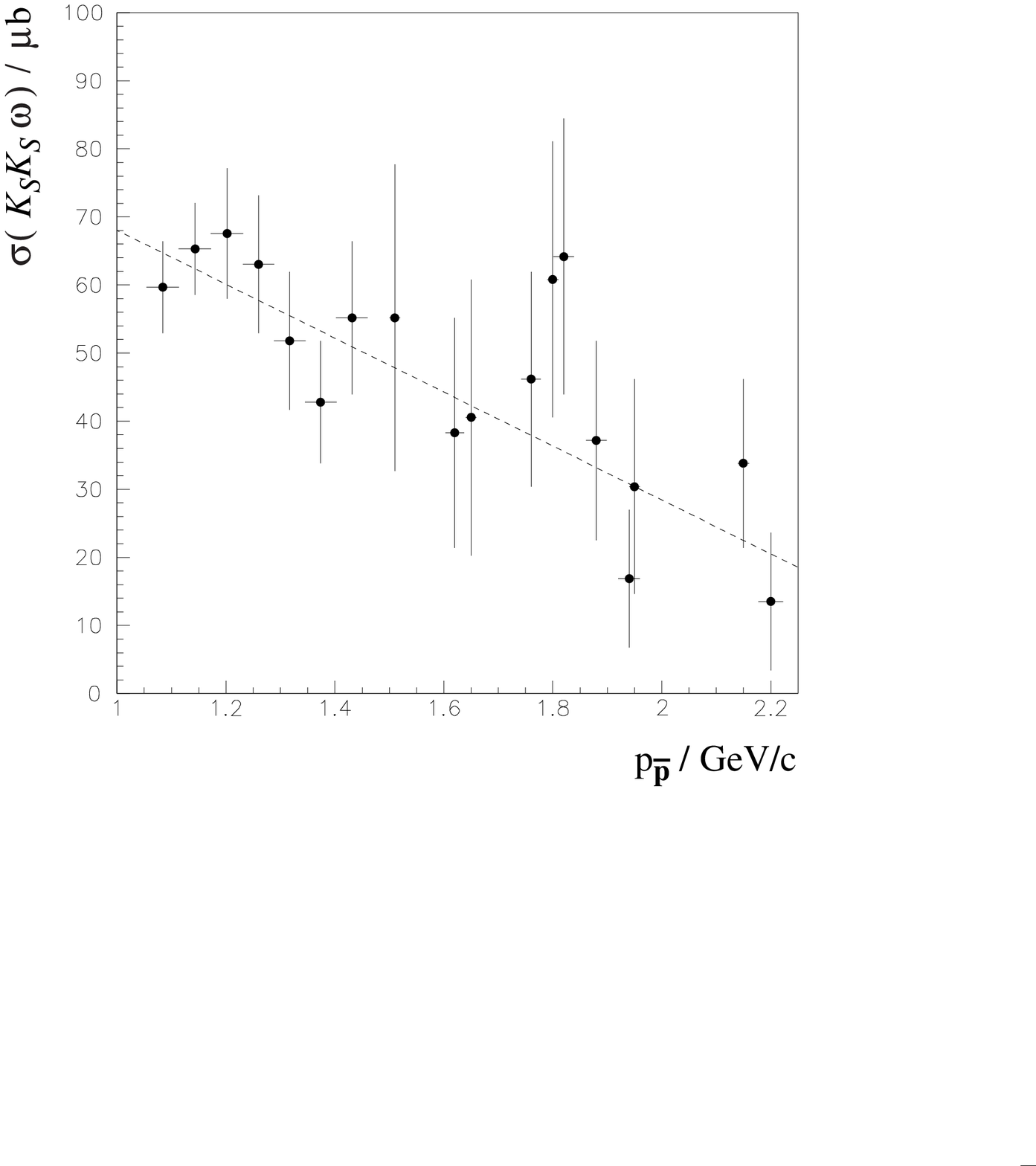}}
 \caption{
   Linear fit to the cross section data for {\ksksom} from
   Refs.~\protect\cite{Chap71,Dona72,Oh73}.
  }
 \label{fig:normal}
\end{center}
\end{figure}

We rely on the known {\ksksom} cross section
to obtain the absolute normalisation of the {\kskseta} cross section.
Data on the {\ksksom} cross section are found in
Refs.~\cite{DAnd68,Chap71,Dona72,Oh73,Oh273,Gang81}
and summarised in Ref.~\cite{crosssec}.
It is worth noting
that all the data, except those of Ganguli {\it et al.}~\cite{Gang81},
have to be corrected for unseen decay modes before the total cross
section can be obtained.
For this analysis, we confined ourselves to the data in
the {\pbar} momentum range from 1.0 to 2.2~GeV/c.
The cross sections given in Ref.~\cite{Oh273}
have not been included since they
are averages of values already given in Ref.~\cite{Oh73}.
The remaining 18 data points shown in Fig.~\ref{fig:normal}
have been corrected for the unseen $\omega$ decay modes
using the branching ratio
BR$(\omega\rightarrow\pi^+ \pi^- \pi^0) = 0.89$~\cite{pdg2000}.
A linear fit is sufficient to describe the data
in this region ($\chi^2_{red} = 0.65$). From the fit we obtain
$\sigma \left(  \pbar p \rightarrow \ks \ks \omega  \right)
 = (50.0 \pm 2.5)~\mu$b at $p_{\overline{p}} = 1.455$~GeV/c
and $(39.7 \pm 2.0)~\mu$b at $p_{\overline{p}} = 1.715$~GeV/c.

%%%%%%%%%%%%%%%%%%%%%%%%%%%%%%%%%%%%%%%%%%%%%%%%%%%%%%%%%%%%%%%%%%%%%%%%%%%%%%%%
%
%                    omega cross section fit
%
%%%%%%%%%%%%%%%%%%%%%%%%%%%%%%%%%%%%%%%%%%%%%%%%%%%%%%%%%%%%%%%%%%%%%%%%%%%%%%%%
%%%%%%%%%%%%%%%%%%%%%%%%%%%%%%%%%%%%%%%%%%%%%%%%%%%%%%%%%%%%%%%%%%%%%%%%%%%%%%%%

%%%%%%%%%%%%%%%%%%%%%%%%%
%\subsection{Cross section}
%%%%%%%%%%%%%%%%%%%%%%%%%

The cross section for the {\kskseta} reaction is finally deduced using the
relation
\begin{equation}
  \sigma \left(  \pbar p \rightarrow \ks \ks \eta  \right)
  \ = \
  \frac{\mbox{number of } \eta\   \mbox{events}}
       {\mbox{number of } \omega\ \mbox{events}}
  \times \frac{\epsilon(\omega)}{\epsilon(\eta)}
  \times f_{neut}\times
  \sigma \left(  \pbar p \rightarrow \ks \ks \omega  \right),
 \label{eq:cross}
\end{equation}
where $\epsilon(\omega)$ and $\epsilon(\eta)$ denote
the acceptances of the corresponding reactions and $f_{neut}$ is
the relative suppression of $\omega$ events compared to $\eta$ events
due to the different neutral branching ratios.
The branching ratio
for neutral decays is $(9.0\pm0.7)$\% for the $\omega$
and $(71.6\pm0.4)$\% for the $\eta$ \cite{pdg2000}
leading to $f_{neut} = 0.126 \pm 0.010$.

Inserting the numbers quoted above into Eq.~\ref{eq:cross}
leads to cross sections for {\kskseta} of
$(3.44 \pm 0.69 \pm 0.50)~\mu$b at $1.455\pm0.020$~GeV/c and
$(2.47 \pm 0.71 \pm 0.36)~\mu$b at $1.715\pm0.060$~GeV/c,
where the first uncertainty is statistical and the second systematical.

Many systematic
uncertainties cancel when forming  ratios as in Eq.~\ref{eq:cross}.
As a cross check of the method, we derived also
the {\kskspi} cross section in a similar way.
We obtained the number of
$\ks \ks \pi^0$ events from the missing mass plot.
The sample also contains some {\ksks} events lying under
the $\pi^0$ peak.
The cross section for this reaction is small,
about 2 $\mu$b~\cite{ksks,crosssec}, and this, together with the
lower acceptance, allows us to neglect its contribution within the precision
of the present experiment.

\begin{figure}[htp!]
\vspace{-0.3cm}
\begin{center}
 \resizebox{0.52\textwidth}{!}{
 \includegraphics*[2.5cm,7.0cm][18.0cm,24.0cm]{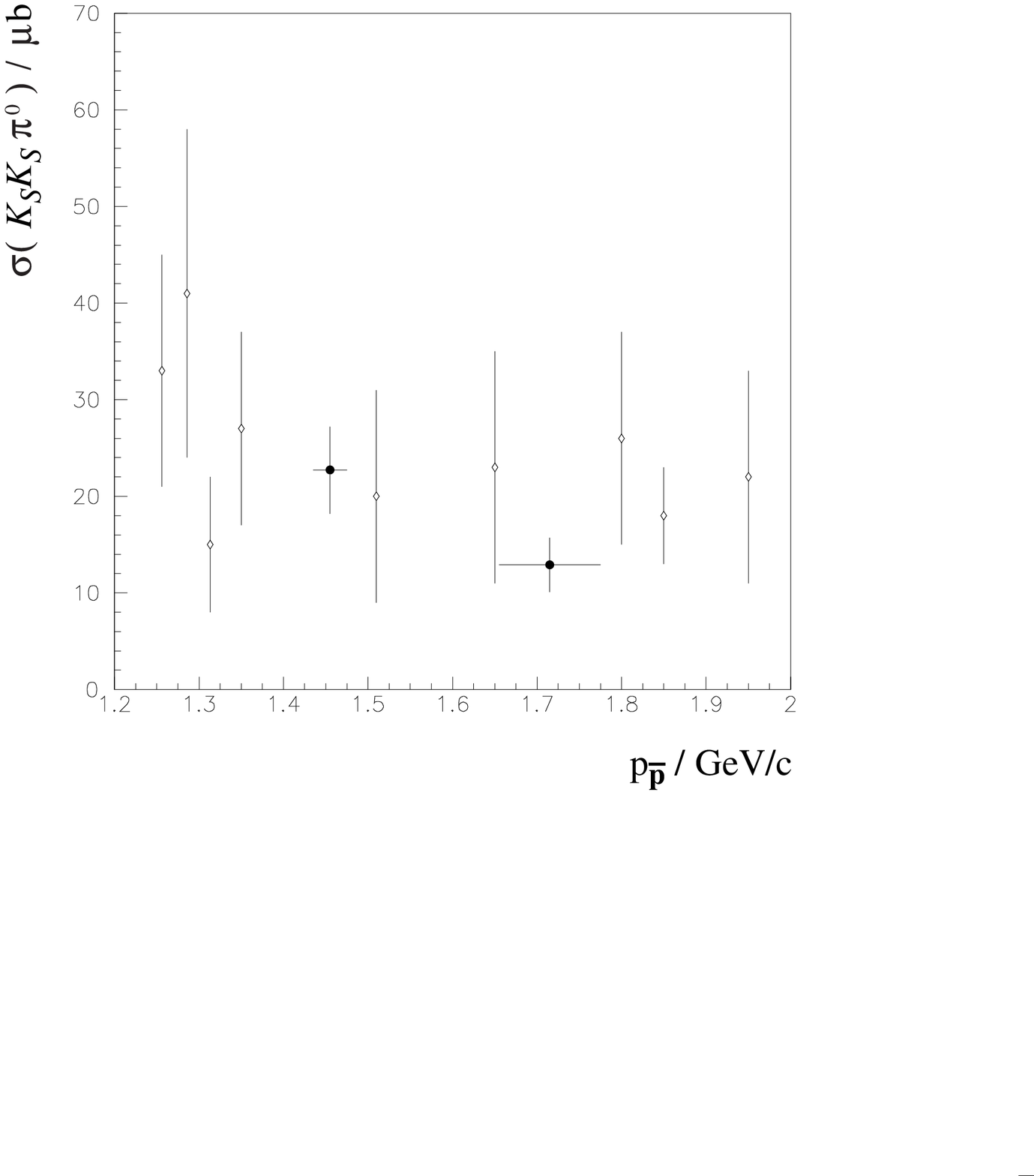}}
 \caption{
   Comparison of our results for
   the {\kskspi} cross section (solid points) with
   values given in Ref.~\protect\cite{crosssec}.
   For the latter values, the uncertainties in the momenta are not shown.
  }
 \label{fig:pi0cross}
\end{center}
\end{figure}

The ratios of the number of $\pi^0$ to $\omega$ events are
$6.23 \pm 0.86 \pm 0.62$ (1.455~GeV/c) and
$3.85 \pm 0.61 \pm 0.38$ (1.715~GeV/c).
The relative efficiency for measuring the  {\ksksom} and {\kskspi} reactions
becomes
$0.80 \pm 0.04$ and $0.93 \pm 0.05$ at the two momenta and
$f_{neut} = 0.091 \pm 0.007$.
With these values,
we find $(22.7 \pm 3.1 \pm 3.3)~\mu$b
at $1.455\pm0.020$~GeV/c and $(12.9 \pm 2.0 \pm 1.9)~\mu$b
at $1.715\pm0.060$~GeV/c for the {\kskspi} cross section.
The good agreement between these results and the
published data~\cite{crosssec} shown in Fig.~\ref{fig:pi0cross} gives us
further confidence in the analysis.

%%%%%%%%%%%%%%%%%%%%%%%%%%%%%%%%%%%%%%%%%%%%%%%%%%%%%%%%%%%%%%%%%%%%%%%%%%%%%%%%
%
%                      Ks Ks pi0 cross section
%
%%%%%%%%%%%%%%%%%%%%%%%%%%%%%%%%%%%%%%%%%%%%%%%%%%%%%%%%%%%%%%%%%%%%%%%%%%%%%%%%
%%%%%%%%%%%%%%%%%%%%%%%%%%%%%%%%%%%%%%%%%%%%%%%%%%%%%%%%%%%%%%%%%%%%%%%%%%%%%%%%

The only published data on {\kskseta} is the branching ratio
for annihilation at rest,
BR$({\kskseta})=(0.25\pm0.04)\times10^{-3}$~\cite{Bara67}.
The corresponding value for the $\ks \ks \omega$ annihilation channel is
BR$({\ksksom})=(1.16\pm0.06)\times10^{-3}$~\cite{Bara67,Bizz71}.
In order to compare these data with the results of our measurement,
we extrapolated the ratio of the $\ks\ks\eta$ and $\ks\ks\omega$ cross
sections
from the threshold value to the momentum range between 1.2 and 2.0~GeV/c.
To do this, we retained only phase space factors,
assuming that the ratio of the matrix elements for both
transitions is constant.
Such considerations predict cross section ratios of 
$0.043 \pm 0.007$ at 1.455~GeV/c and
$0.039 \pm 0.006$ at 1.715~GeV/c.
Our data points lie within one standard deviation above this estimate.

In conclusion, we have made the first measurement of the cross section
for the {\kskseta} reaction in flight in two regions of $\overline{p}$
momenta,
around 1.455 and 1.715~GeV/c.
The magnitude of the cross section is in reasonable agreement with
an extrapolation from the known
{\kskseta} and {\ksksom} branching ratios at rest
using phase space arguments.

We hope that these results will stimulate further theoretical 
efforts to shed light on the process of strangeness production.

%%%%%%%%%%%%%%%%%%%%%%%%%%%%%%%%%%%%%%%%%%%%%%%%%%%%%%%%%%%%%%%%%%%%%%%%%%%%%%%%

%%%%%%%%%%%%%%%%%%%%%%%%%%%%%%%%%%%%%%%%%%%%%%%%%%%%%%%%%%%%%%%%%%%%%%%%%%%%%%%%
%
\ack
%                         ACKNOWLEDGEMENTS
%
%%%%%%%%%%%%%%%%%%%%%%%%%%%%%%%%%%%%%%%%%%%%%%%%%%%%%%%%%%%%%%%%%%%%%%%%%%%%%%%%

%%%%%%%%%%%%%%%%%%%%%%%%%%%%%%%%%%%%%%%%%%%%%%%%%%%%%%%%%%%%%%%%%%%%%%%%%%%%%%%%

The PS185 team wants to thank the LEAR accelerator team for providing
excellent experimental conditions.
We would also like to thank Professor Colin Wilkin for a very helpful
critique of our manuscript.
We also gratefully acknowledge the financial and material support from
the German Bundesministerium f{\"u}r Bildung und Forschung,
the Swedish Natural Science Research Council,
the United States Department of Energy and
the United States Science Foundation.
This work is based in part on the dissertation of S.~Pomp submitted to
the Uppsala University in partial fulfillment of the requirements
for the PhD degree.

%%%%%%%%%%%%%%%%%%%%%%%%%%%%%%%%%%%%%%%%%%%%%%%%%%%%%%%%%%%%%%%%%%%%%%%%%%%%%%%%

%%%%%%%%%%%%%%%%%%%%%%%%%%%%%%%%%%%%%%%%%%%%%%%%%%%%%%%%%%%%%%%%%%%%%%%%%%%%%%%%

%

%%%%%%%%%%%%%%%%%%%%%%%%%%%%%%%%%%%%%%%%%%%%%%%%%%%%%%%%%%%%%%%%%%%%%%%%%%%%%%%%

%%%%%%%%%%%%%%%%%%%%%%%%%%%%%%%%%%%%%%%%%%%%%%%%%%%%%%%%%%%%%%%%%%%%%%%%%%%%%%%%
%
%
\end{document}